\documentclass[prl,twocolumn,showpacs,superscriptaddress,preprintnumbers,amssymb]{revtex4}
\usepackage{graphicx}
\usepackage{dcolumn}
\usepackage{bm}
\usepackage{wasysym}
\usepackage{psfrag}
\input epsf

\newcommand{\beq}{\begin{equation}}
\newcommand{\eeq}{\end{equation}}
\newcommand{\beqn}{\begin{eqnarray}}
\newcommand{\eeqn}{\end{eqnarray}}

\begin{document}
\title{Entanglement entropy of 2D conformal quantum critical points: \\
hearing the shape of a quantum drum 
}
\author{Eduardo Fradkin}
\affiliation{Department of Physics, University of Illinois, 1110 West Green Street, Urbana, Illinois 61801-3080}
\author{Joel E.~Moore}
\affiliation{Department of Physics, University of California, Berkeley, CA 94720}
\affiliation{Materials Sciences Division, Lawrence Berkeley National Laboratory, Berkeley, CA 94720}
\date{\today}
\begin{abstract}
The entanglement entropy of a pure quantum state of a bipartite system $A \cup B$ is defined as the von Neumann entropy of the reduced density matrix obtained by tracing over one of the two parts.  Critical ground states of local Hamiltonians in one dimension have entanglement that diverges logarithmically in the subsystem size, with a universal coefficient that for conformally invariant critical points is related to the central charge of the conformal field theory.  We find the entanglement entropy for a standard class of $z=2$ quantum critical points in two spatial dimensions with scale invariant ground state wave functions: in addition to a nonuniversal ``area law'' contribution proportional to the size of the $AB$ boundary, there is generically a universal logarithmically divergent correction.  This logarithmic term is completely determined by the geometry of the partition into subsystems and the central charge of the field theory that describes the equal-time correlations of the critical wavefunction.  
\end{abstract}

\pacs{03.67.Mn, 11.25.Hf}

\maketitle

A major challenge in the theory of quantum critical phenomena is to understand those aspects that do not appear in the classical theory.  The entanglement entropy at criticality is an example: a pure quantum state of a bipartite system $A \cup B$ can become a mixed state, with an associated entropy, when restricted to subsystem $A$ or $B$.  Critical ground states in one dimension are known to have entanglement entropy that diverges with subsystem size for several types of pure~\cite{holzhey,Vidal03,Calabrese04,Kopp06} and disordered~\cite{refaelmoore,laflorencie,santachiara} quantum critical points, with a coefficient determined by the central charge in the pure case.  Hence, in one dimension, entanglement entropy gives a definition of universal critical entropy for quantum critical points that is consistent with the conventional definition (central charge) in the conformally invariant case.

In higher dimensions, results have been obtained recently for gapless free fermions~\cite{klich,wolf}, and a generic scaling form is conjectured in~\cite{Calabrese04}.  A standard expectation~\cite{srednicki}, which is known to be violated for free fermions, is of an ``area law'': entanglement entropy scales as the area of the boundary between the subsystems.  For example, if only one subsystem is finite, the area law predicts that entanglement entropy scales as $L^{d-1}$ in $d$ spatial dimensions, where $L$ is the linear size of the finite subsystem.  Entanglement entropy in gapped phases in two dimensions satisfies an area law but has subleading terms that probe topological order~\cite{Levin05,Kitaev06b}: adding and subtracting properly chosen regions cancels the leading area term and leaves a term reflecting the ground-state degeneracy on topologically nontrivial manifolds.

An additional motivation for recent studies of entanglement entropy in many-body physics is as a route to better numerical algorithms for finding ground states: knowing that ground states of local Hamiltonians have much less entanglement, even at criticality, then generic quantum states both explains the success of the ``density-matrix renormalization group'' algorithm in one dimension~\cite{white,ostlundrommer,vidalrg} and motivates recent proposals for analogous numerical methods in higher dimensions~\cite{Verstraete06}.

This paper obtains the entanglement entropy for the class of ``conformal'' two-dimensional quantum critical points\cite{Ardonne04} that includes such standard examples as the quantum dimer model~\cite{Rokhsar88,moessner02a}, the quantum eight-vertex model~\cite{Ardonne04}.  These quantum critical points were first introduced because perturbing away from these solvable critical points can yield topologically ordered phases in lattice models~\cite{moessnersondhi}.  The same method can be used to obtain the entanglement entropy for related noncritical wavefunctions.  These critical points have dynamic critical exponent $z=2$ and equal-time correlation functions given by a two-dimensional conformal field theory (CFT).  For the entanglement entropy created by partitioning of a 2D manifold into regions $A$ and $B$, we find a universal subleading logarithmic contribution in addition to the ``area law'' nonuniversal contribution proportional to the linear size $L$ of the boundary:
\beq
S = 2 f_s (L/a) + \alpha c  \log (L/a) + O(1).
\eeq
Here $c$ is the central charge of the CFT, $a$ is the ultraviolet cutoff, and the coefficient $\alpha$ is determined by geometric properties of the partition.  The area law coefficient $f_s$ is interpreted below as a boundary free energy.  As an example of the geometric dependence of $\alpha$, if $A$ is a rectangle surrounded by $B$, $\alpha=- {1 \over 9}$, while if $A$ has a smooth boundary surrounded by $B$, $\alpha = 0$.  This contribution for the case of a free scalar field originates in the logarithmic contribution that appears in the classical problem of the determinant of the Laplacian operator on a 2D manifold~\cite{Kac66,McKean67} (``hearing the shape of a drum''), which was extended to a general CFT by Cardy and Peschel~\cite{Cardy88}.

The entanglement entropy of a pure state of a bipartite system is defined as the von Neumann entropy of the reduced density matrix for either subsystem:
\beq
S = - {\rm Tr}\,\rho_A \log \rho_A = - {\rm Tr}\,\rho_B \log \rho_B.
\eeq
For conformal quantum critical points, the Hilbert space has an orthonormal basis of states $|\{ \phi \} \rangle$ indexed by classical configurations $\{ \phi \}$, and the ground state $|\psi_0\rangle$ of the bipartite system is determined by a CFT action $S$:
\beq
|\psi_0\rangle = {1 \over \sqrt{Z_c}} \int\,(d\phi)\,e^{-S(\{ \phi \}) /2} |\{\phi\}\rangle.
\eeq
Here $Z_c$ is the partition function $\int\,(d\phi)\,e^{-S(\{\phi\})}$, and expectation values in this state reproduce CFT correlators.

The next step is to obtain traces of powers of the reduced density matrix in order to find the entanglement entropy using the replica trick
\beq
-{\rm Tr}\,\rho \log \rho = - \lim_{n \rightarrow 1} {\partial \over \partial n} {\rm Tr}\, \rho^n.
\eeq  First consider the reduced density matrix $\rho_A$.  The element of the reduced density matrix between internal configurations $|\{\phi^A_1\}\rangle$ and $|\{\phi^A_2\}\rangle$ is, after introducing a normalization factor to ensure ${\rm Tr}\,\rho_A = 1$,
\begin{widetext}
\beqn
\langle \{\phi^A_1\}|\rho_A | \{\phi^A_2\} \rangle &=& {\rm Tr}_{\phi^B}\,(\langle \{\phi^A_1\}| \otimes \langle \{\phi^B \}| \psi_0\rangle \langle \psi_0 (| \{\phi^A_2\} \rangle \otimes |\{\phi^B\} \rangle \cr
&=& {1 \over Z_c} \int (d\phi^B) e^{-(S^A(\phi^A_1)/2 + S^A(\phi^A_2)/2 + S^\partial(\phi^A_1,\phi^B)/2 + S^\partial(\phi^A_2,\phi^B)/2 + S^B(\phi^B))}.
\eeqn
\end{widetext}
Here the action has been divided into regions $A$, $B$, and the boundary $\partial$, where the last takes into account contributions mixing the $A$ and $B$ degrees of freedom (e.g., couplings of spins across the boundary in a lattice model).

Higher powers of the density matrix need not trace to unity: ${\rm Tr}\,\rho^n_A$ is now a sum over $n$ configurations defined in $A$ and $n$ configurations defined in $B$.  The key is to keep track of how these different configurations are stitched together at the boundary by the terms $S^\partial$ that link $A$ and $B$: $\{\phi^A_i\}$ is linked to $\{\phi^B_i\}$ as well as $\{\phi^B_{i+1}\}$ for $i=1, \ldots, n-1$, and $\{\phi^A_n\}$ is linked to $\{\phi^B_n\}$ and $\{\phi^B_1\}$.  This is normalized through division by $(Z_c)^n$, which can be thought of again as $n$ copies of $A$ and $B$ configurations, but with $\{\phi^A_i\}$ linked only to $\{\phi^B_i\}$.

In the continuum limit, the boundary terms impose continuity of the fields in the CFT because strong local fluctuations are penalized in the action.  Since each $A$ field is linked to the $B$ field of the same index, we can define global configurations $\{\phi_i \}$ in both the numerator and denominator, but in the numerator, the links between configuration $i$ and configuration $i \pm 1$ require that all $n$ configurations agree on the boundary, while this requirement is absent in the denominator. 
Schematically, \beq
{\rm Tr}\,\rho_A^n = {Z(\hbox{$n$ configurations agreeing on the boundary}) \over
Z(\hbox{$n$ independent configurations})}.
\eeq
Note that this is symmetric with respect to $A$ and $B$: ${\rm Tr}\,\rho_A^n = {\rm Tr} \,\rho_B^n$ as known from Schmidt decomposition.

If ${\rm Tr}\,\rho_A^n$ is known for all integer $n$, then assuming an analytic continuation, the entropy is obtained as
\beq
S = - {\partial {\rm Tr}\,\rho_A^n \over \partial n}\Big|_{n=1}.
\eeq
The above expression for ${\rm Tr}\,\rho_A^n$ can be put in a form that simplifies taking this derivative: for an explicit realization, consider the case of a free scalar field.  Then the condition that $n$ scalar fields $\phi_i$ agree with each other on the boundary can be satisfied by forming $n-1$ linear combinations ${1 \over \sqrt{2}} (\phi_i - \phi_{i+1})$, which vanish at the boundary (i.e., satisfy Dirichlet boundary conditions), plus one linear combination ${1 \over \sqrt{n}} \sum_{i=1,\ldots,n} \phi_i$ that has no restriction at the boundary (i.e., is a free field on $A \cup B$).

More generally, for any CFT there exists a conformal boundary condition that generalizes the notion of the Dirichlet boundary condition in the free scalar case.  Thus, in terms of the partition functions $Z_D$, for a field in the whole system $A \cup B$ that vanishes at the boundary, and $Z_F$, for a free field in the whole system,
\beq
{\rm Tr}\,\rho_A^n = {Z_D^{n-1} Z_F \over {Z_F}^n} = \left({Z_D \over Z_F}\right)^{n-1}
\eeq
and therefore
\beq
S = -\log {Z_D \over Z_F} = - \log {Z_D^A Z_D^B \over Z_F}.
\eeq
In the last equality,  the Dirichlet boundary condition at the boundary was used to split the partition function into contributions from $A$ and $B$, each including the boundary with Dirichlet boundary conditions.  Finally, the entanglement entropy for a general conformal quantum critical point is just the dimensionless free energy difference induced by the partition in the associated CFT:
\beq
S = F_A + F_B - F_{A \cup B}.
\label{composition}
\eeq

Explicit results for the entanglement entropy can now be found using results on the free energy corrections in conformal field theory.  If $A \cup B$ is the plane, $A$ and $B$ are connected (which requires that one be simply connected), and the boundary is smooth, then the free energy result of Cardy and Peschel~\cite{Cardy88} can be applied: using region $A$ as an example and supposing that it is finite,
\beq
F_A = f_b |A| + f_s L - {c \chi \over 6} \log L + O(1)
\label{FA}
\eeq
where $f_b$ and $f_s$ are the bulk and surface partial densities, $|A|$ is the area, $L$ is the perimeter, and $\chi$ is the Euler characteristic of the manifold
\beq
\chi = 2 - 2 g - b.
\label{euler}
\eeq
Here $g$ is the number of handles and $b$ the number of boundaries of the manifold.   As an example, $\chi = 1$ for a planar simply connected region with boundary, such as a disk.  In (\ref{FA}),  the cutoff is now incorporated into the definitions of $f_b$ and $f_s$: note that the coefficient of the logarithm is cutoff-independent, as changing the cutoff gives only an additive constant from this term.  

In the expression for $S$, the bulk energy terms cancel, and since $A \cup B$ is the plane and has no boundary, the linear in $L$ term is just twice the dimensionless boundary energy in the CFT with the same regularization.  Note that the logarithmic term differs from the leading (boundary) term in that it is cutoff-independent and hence potentially universal.  The logarithmic contribution is determined for a smooth boundary by the change in Euler characteristic:
\beq
S = 2 f_s L - {c \over 6} (\chi_A + \chi_B - \chi_{A \cup B}) \log L.
\label{composition-entropy}
\eeq
We note immediately from this formula that if the boundary between $A$ and $B$ is a smooth closed curve in the interior of $A \cup B$, there is no logarithmic contribution because the total Euler characteristic is conserved.  Although this cancellation is quite specific to a smooth closed boundary, it shows that the existence of logarithmic contributions to $F_A$ and $F_B$ (unless they have zero Euler characteristic) does not imply a logarithmic entanglement entropy. 

There are two simple mechanisms that generate a logarithmic contribution.  The first is that the boundary between $A$ and $B$ may intersect the boundary of $A \cup B$, as for example if a disk is sliced into two halves, which modifies the total Euler characteristic.  The second is that the boundary between $A$ and $B$ may have corners.~\cite{footnote-conical}  The Cardy-Peschel formula argument given here is valid for boundaries that are smooth except for a finite number of sharp corners; fractal boundaries, for example, could give rise to different subleading corrections.
Henceforth we write $S_{\log}$ for the possible logarithmic term in the entropy.  Three examples are shown in Fig. 1.

We first treat the case of sharp corners.  A sharp corner with interior angle $\gamma$, $0 < \gamma < 2 \pi$, gives a logarithmic contribution to the free energy~\cite{McKean67,Cardy88}
\beq
\Delta F = {c \gamma \over 24 \pi} \left(1 - {\pi^2 \over \gamma^2} \right) \log L
\label{anglecont}
\eeq
According to the formula (\ref{anglecont}), the contribution from a polygon approaches that of a smooth curve as all angles approach $\pi$.  However, sharper angles contribute more to the logarithmic correction than would be predicted by the Gauss-Bonnet formula.  Contributions from the angles $\gamma$ and $2 \pi - \gamma$ only cancel to linear order near $\gamma=\pi$, so that sharp angles on the $AB$ boundary give a net contribution.

As an example, summing the interior and exterior contributions for the four corners of a rectangle $A$ surrounded by $B$ gives
\beq
S_{\log} = c \left({5 \over 36}- {1 \over 4}\right ) \log L = - {c \over 9} \log L.
\eeq
Note that a negative or positive sign for the logarithmic part of the entropy is physically permissible as a subleading correction to the (positive) boundary entropy.

\begin{figure}[h!]
\epsfxsize = 3.0in
\epsffile{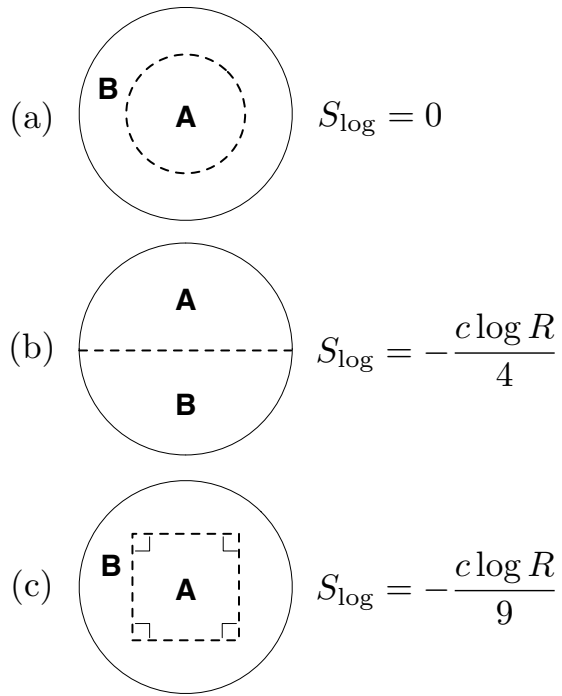} \caption{The logarithmic contribution to entanglement entropy for three partitions of a disk.  Partitions (a) and (c) do not change the total Euler characteristic, while partition (b) does.  Note that it is the partition length scale, which may be finite even if $A$ or $B$ is infinite, that sets $R$ in the logarithm.}
\end{figure}

Another source of logarithmic contributions is a smooth boundary that {\em disconnects the original system}.  As an example, consider the disk of radius $L$ cut into two pieces by a diameter.  The resulting two half-disks have two sharp angles of ${\pi \over 2}$ each, formed by the intersection of the diameter with the circumference: the resulting logarithmic contribution is
\beq
S_{\log} = 4 {c (\pi / 2) \over 24 \pi} (1 - 4) \log L = - {c \over 4} \log L.
\eeq

The quantum critical points studied in this paper can, as in the quantum dimer model and its generalizations (both Abelian and non-Abelian)\cite{Ardonne04,Fendley05,Seidel05}, lie next to topologically ordered phases with subleading $O(1)$ terms in their entanglement entropy in addition to an area law:~\cite{Kitaev06b,Levin05} for some nonuniversal constant $a$ and a universal topological number $\gamma$,
\beq
S_{\rm topo} = a L + \gamma + O(1/L). 
\eeq

Another way to obtain a universal logarithmic term at criticality is by means of a {\em physical process} by which the system is split into two disjoint regions each with a smooth boundary. In this case there is a net change in the Euler characteristic leading to a lowering of the universal term in the entanglement entropy (due to a physical loss of correlations) by $\Delta S=-(c/6) \log L$. Something very similar has been found to happen to the universal constant term $\gamma$ which arises in a {\em topological fluid} (see above), such as in the case of a quantum Hall fluid\cite{Fendley06}.

It is natural to ask whether even at criticality there are topological nondivergent terms in addition to the universal logarithmic divergences studied here.  For the case of the free scalar field, there can be nondivergent terms with a complicated (but scale invariant) dependence on the partition geometry, suggesting that there is no purely topological number\cite{Weisberger87}: e.g., the free energy of the annulus with radii $r_1 < r_2$ contains a term $\log \log (r_2 / r_1)$.  

An important question for future work is to understand explicitly how the entanglement evolves, for a topologically nontrivial partition, as the Hamiltonian is tuned from a critical point to a massive topological phase, as this requires a more accurate calculation than simply cutting off the logarithmic divergences by a correlation length $\xi$ away from criticality. The dependence of the universal logarithmic term on the central charge $c$ of the associated 2D CFT indicates that upon perturbation there should be a downward flow of the coefficient of this universal term.

Although the area law at leading order is consistent with the general scaling ansatz of Calabrese and Cardy\cite{Calabrese04} in $d>1$,
our results show that in $d=2$ there are universal logarithmic contributions to the entanglement entropy at a class of critical points for which the ``area law'' applies.  These universal contributions are determined by the critical bulk theory and the geometric details of the partition into subsystems.  Combined with recent results on subleading corrections to entanglement entropy in massive topological phases~\cite{Levin05,Kitaev06b}, it now appears that entanglement entropy in $d \geq 2$ is considerably richer than the area law might suggest, even when the area law is applicable.

We are grateful to P. Fendley, M. P. A. Fisher, M. Freedman, P. Goldbart, I. Klich, C. Nayak, J. Preskill, G. Refael, M. Srednicki, M. Stone, and F. Verstraete for many discussions. 
This work was supported in part by the NSF grant DMR 0442537 (EF), NSF DMR-0238760 (JEM), and NSF PHY99-07949,  and the hospitality of the Kavli Institute for Theoretical Physics.


\begin{thebibliography}{28}
\bibitem[{foo()}]{footnote-conical}
\bibinfo{howpublished}{We ignore the possibility of conical singularities in
  the metric near the boundary, which can also generate logarithmic
  contributions.}
\end{thebibliography}

\end{document}